\documentclass[useAMS,usenatbib]{mn2e}
\usepackage{times}
\usepackage{graphicx}
\usepackage{amssymb}
\usepackage{amsmath}
\title[Ultra compact dwarfs in Perseus]{Ultra compact dwarfs in the Perseus Cluster: UCD formation via tidal stripping}
\author[S. J. Penny et al.]{Samantha~J.~Penny$^{1,2}$, Duncan A. Forbes$^{3}$, Jay Strader$^{4}$, Christopher Usher$^{3}$ \newauthor  Jean P. Brodie$^{5}$ and Aaron J. Romanowsky$^{5,6}$ \footnotemark[0]\\
$^1$School of Physics, Monash University, Clayton, Victoria 3800, Australia\\ 
$^2$Monash Centre for Astrophysics, Monash University, Clayton, Victoria 3800, Australia\\
$^3$Centre for Astronomy and Supercomputing, Swinburne University, Hawthorn, Victoria 3122, Australia\\
$^4$Department of Physics and Astronomy, Michigan State University, East Lansing, MI 48824, USA\\
$^5$University of California Observatories and Department of Astronomy and Astrophysics, University of California, Santa Cruz, CA 95064, USA\\
$^6$Department of Physics and Astronomy, San Jos\'e State University, One Washington Square, San Jose, CA 95192, USA
}
\begin{document}

\maketitle

\begin{abstract}
We present the results of a Keck/DEIMOS survey of Ultra Compact Dwarfs (UCDs) in the Perseus Cluster core. We confirm cluster membership for 14 UCDs, with radial velocities $\sim5300$~km~s$^{-1}$. Two of these confirmed Perseus UCDs have extremely blue colours ($B-R < 0.6$~mag), reside in star forming filaments surrounding NGC~1275, and have likely formed as massive star clusters in the last $\sim100$~Myr. 
We also measure a central velocity dispersion of a third, UCD13 ($\sigma_{0} = 38 \pm 8$~km~s$^{-1}$), the most extended UCD in our sample. 
We determine it to have radius $R_{e} = 85 \pm 1.1$~pc, a dynamical mass of $(2.3 \pm 0.8)\times10^{8}$~M$_{\sun}$, and a metallicity [Z/H]$ = -0.52^{+0.33}_{-0.29}$~dex. 
UCD13 and the cluster's central galaxy, NGC 1275, have a projected separation of 30 kpc and a radial velocity difference of $\sim20$~km~s$^{-1}$.
Based on its size, red colour, internal velocity dispersion, dynamical mass,  metallicity and proximity to NGC~1275, we argue that UCD13 is  likely the remnant nucleus of a tidally stripped dE, with this progenitor dE having $M_{B} \approx -16$~mag and mass $\sim10^{9}$~M$_{\odot}$. 
\end{abstract}

\begin{keywords} 
galaxies: star clusters: general -- galaxies: clusters: individual: Perseus Cluster -- galaxies: dwarf
\end{keywords} 
\section{Introduction}

Ultra Compact Dwarfs (UCDs) were first identified over a decade
ago (Hilker et al. 1999; Drinkwater et al. 2000), but whether they
should be considered as galaxies or massive star clusters is
still uncertain (e.g. Hasegan et al. 2005; Mieske et al. 2009;
Hilker 2009; Forbes \& Kroupa 2011; Willman \& Strader
2012). 
Brodie et al. (2011) adopted a working definition for UCDs to have half-light sizes of $10~{\rm pc} < R_{h}<100$~pc and $M_{V} < -8$~mag (with no upper limit on luminosity).

The two main formation channels discussed in the literature are
variations of merged star clusters \citep{1998MNRAS.300..200K,2002MNRAS.330..642F,2011A&A...529A.138B} or stripped dwarf galaxies \citep{1994ApJ...431..634B,2001ApJ...552L.105B}. 
In this sense their current size and luminosity are thought to be the result of a `bottom-up'
growth from lower masses, or a `top-down' evolution from higher
masses.  
Support for the idea of both formation channels for UCDs comes 
from the results of  Norris \& Kannappan (2011). They found many UCDs to be consistent
with high mass globular clusters. However, they also reported one UCD around
NGC 4546 that was in a counter-rotating orbit with respect to the
host galaxy and its globular cluster system. Furthermore, they calculated that the UCD had a very
short dynamical friction timescale. This strongly suggested that
it had been recently captured and was not part of the galaxy's original
GC system. 

Evidence for UCDs forming as massive star clusters in the nearby Universe was presented by Penny et al. (2012), who found young, blue candidate UCDs with ($B-R$)$_{0}<0.6$~mag in the Perseus Cluster associated with gas filaments around the cluster's central galaxy NGC 1275. 
NGC1275 therefore provides a location in which to study the formation of massive star clusters and proto-UCDs in the nearby Universe. 
If these massive blue star clusters are able to remain bound against the cluster tidal potential, they will be seen as typical red UCDs once their stellar populations fade in a timescale of $\sim500$~Myr.  

Using a state-of-the-art compilation of pressure-supported
systems with old stellar populations, Brodie et al. (2011) 
found that UCDs had a tight colour-magnitude relation that was
different from normal globular clusters. This, and kinematic
differences, argued against the merged star cluster scenario. 
However, they also found that UCDs do not have a 
well-defined size-luminosity distribution as
might have been expected in the stripping scenario \citep{2013MNRAS.433.1997P}. A further
expectation of the stripping scenario is that the remnants come
from a range of progenitor masses. Hence we would expect a
continuity of objects between the most massive UCDs and compact
ellipticals (cEs), the latter of which are thought to be the 
tidally stripped remnant of a normal galaxy (Faber 1973). Such
intermediate mass objects are too massive to be explained as merged star clusters.

In the compilation of Brodie et al. (2011), their figure 8 shows
very few objects with confirmed sizes and luminosities
intermediate between those of UCDs and cEs. One such object is
M59cO highlighted by Chilingarian \& Mamon (2008) as a ``missing
link'' between UCDs and cEs. This object, with a confirmed velocity
placing it in the Virgo cluster, has  $M_{B} =
-12.34$~mag ($M_V$ $\sim -13.0$~mag). 
A dual Sersic fit to the surface brightness profile
gave a `nucleus' of half-light radius $R_{e}=13$~pc with an extended 
component of $R_{e}=50$~pc. The object was also found to be old and relatively metal-rich. 
Another extreme UCD in the Virgo Cluster, M60-UCD1 \citep{2013ApJ...775L...6S} with $M_{B} = -13.24$~mag and a dynamical mass $M_{dyn} = (2.0 \pm 0.3) \times 10^{8}$~M$_{\sun}$ is the densest galaxy identified to date, and occupies the region of size-magnitude space between the most massive globular clusters and the faintest compact elliptical galaxies. 

In this work we present the results of a Keck DEIMOS follow-up study of UCD candidates in the nearby Perseus Cluster ($D=71$~Mpc) identified in \citet{Penny12} out to 250~kpc from the cluster centre. 
The Brightest Cluster Galaxy (BCG) is the unusual NGC 1275, which is actively accreting/disrupting satellite galaxies at the present time, making Perseus an ideal environment in which to study the formation of UCDs as disrupted dEs vs. massive star clusters. The focus on this paper is the origin of UCDs in clusters- are they massive star clusters or the cores of tidally stripped galaxies? In particular we examine on the origin of the largest Perseus UCD, UCD13, via its size, kinematics, cluster location and stellar population.

This paper is organised as follows. In Section~\ref{sec:data}, we describe our observations and their reduction. In Section~\ref{perucd} we briefly discuss the properties of the confirmed Perseus Cluster members, including the size luminosity distribution for UCDs.  
An analysis of UCD13 is presented in Section~\ref{ucd13}, with a tidal stripping formation scenario for UCD13 examined in Section~\ref{sec:ucd13}. 
A star cluster origin for UCD13 is discussed in Section~\ref{starcluster}.
We discuss our results in Section~\ref{sec:discuss}, and present our conclusions in Section~\ref{sec:conclude}.

\section{Data}
\label{sec:data}

\subsection{HST ACS imaging}

In \citet{Penny12}, from HST ACS WFC imaging we identified 84 candidate UCDs in the Perseus Cluster core using two fields in the $F475W$ ($B$) and  $F625W$ ($R$) bands, along with five fields in the $F555W$ ($V$)  and $F814W$ ($I$) bands. 
These UCD candidates were selected based on their sizes alone ($10~{\rm pc} < R_{e} < 150~{\rm pc}$), with no colour or luminosity cuts placed on their selection. 
Instead, all candidate UCDs were identified to have $S/N > 40$ to allow for an accurate determination of their size via \textsc{ishape} \citep{Larsen99}, along with smooth, round shapes with no evidence for substructure. 
All UCD candidates identified using this method were brighter than $M_{R} = -10.5$~mag.
The reduction and analysis of the $HST$ data are described in \citet{Penny12}. 
The sizes, magnitudes, and colours of all Perseus UCDs presented in this paper are taken from \citet{Penny12}, except  for the size of UCD13 which we remeasure here.  These UCD properties are shown in Table~\ref{ucddat}.

\subsection{Keck DEIMOS Spectroscopy} 
\label{sec:spec}

\begin{figure}
\includegraphics[width=0.48\textwidth]{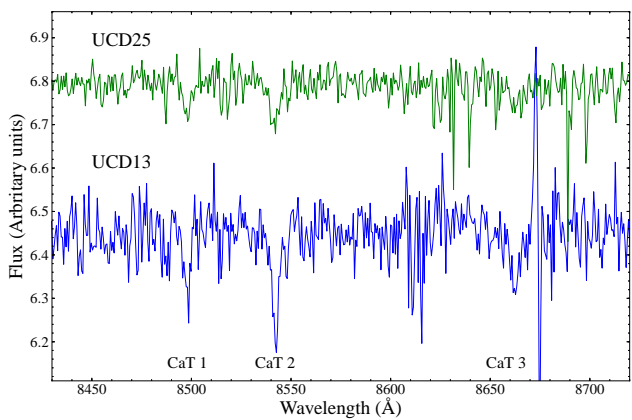}
\caption{Example Keck-\textsc{deimos} spectra for two confirmed Perseus UCDs (UCD13 and UCD25).  The spectra are corrected to rest wavelength using the velocities shown in Table~\ref{ucddat}, and trimmed to show the CaT region, the primary feature we use for red-shifting in this work. An arbritrary offset has been applied to the fluxes of the two spectra. The spectrum for UCD13 has a S/N sufficient for the determination of its internal velocity dispersion. It can be seen that the CaT~3 line is affected by sky lines, however the CaT~1 and CaT~2 lines are uncontaminated. \label{twospecs}}
\end{figure}

We obtained spectra of UCD candidates in the Perseus Cluster core from multi-object \textsc{deimos} spectroscopy on the Keck \textsc{ii} telescope. 
The data were obtained on 2012 February 20th. 
The exposure time was 4800~s, split into four sub-exposures. 
We observed the mask using the 900~lines~mm$^{-1}$ grating with $1''$ slits, providing a spectral resolution ${\rm R}\approx4000$. 
We obtain a velocity resolution of 16~km~s$^{-1}$~pixel$^{-1}$.
These observations provide spectral coverage from $5750$~{\AA} to 7250~{\AA} for the blue chip, and 7500~{\AA} to 9200~{\AA} for the red chip, covering both the H$\alpha$ and Calcium \textsc{ii} triplet (CaT) absorption features at the redshift of Perseus ($v = 5366$~km~s$^{-1}$, Struble \& Rood 1999). 

We targeted 37~UCD candidates in the Perseus Cluster core for spectroscopy. The \textsc{deimos} data were reduced using the \textsc{idl spec2d} pipeline. 
The data were flatfielded, wavelength calibrated using ArKrNeXe arcs, and sky-subtracted, and 14 UCD candidates emerged with sufficient signal-to-noise to determine reliable redshifts. Radial velocities were measured using the \textsc{iraf} procedure \textsc{fxcor}. Example spectra for two UCDs (UCD13 and UCD25) are shown in Figure~\ref{twospecs}. It can be seen that the CaT~3 line for UCD13 is affected by sky lines, so cannot be used in the velocity dispersion determination as described in Section~\ref{sec:cE}. 

The UCD spectra were matched to Vazdekis (2003) model spectra with a range of ages and metallicities, covering both the H$\alpha$ and CaT regions. 
The CaT feature is preferentially used for determining redshifts. For two UCDs with bright H$\alpha$ emission (UCD17 and UCD29), the redshifts were determined manually based on the peak of the H$\alpha$ emission, and confirmed using the CaT absorption features.   

\begin{table*}
\caption{Perseus Cluster UCDs with spectroscopically confirmed cluster membership. Lower and upper limits are presented for the size of UCD9 (see the explanatory text in Section~\ref{sec:spec}). Apparent magnitudes, colours, and effective radii are taken from Penny et al. (2012). The effective radii presented here are determined from King profiles. The radial velocities are obtained from \textsc{keck deimos} spectroscopy. The spectral feature used for the radial velocity determination is also shown. Magnitudes and colours for UCDs 1 through 40 are in $B$ and $R$, and those for UCD60 and UCD65 are in $V$ and $I$ reflecting the available $HST$ imaging.} 
\begin{center}
\begin{tabular}{lccccc}
\hline
Object & R$_{e}$ & R & (B--R)$_{0}$ & $v$ & line used \\
& (pc) & (mag)& (mag)& (km~s$^{-1}$) & \\
\hline
UCD1    & 11.7 & 22.81 $\pm$ 0.02 & 1.18 $\pm$ 0.05 & 5717 $\pm$ 10.9  & CaT\\
UCD2    & 26.9 & 22.90 $\pm$ 0.03 & 1.44 $\pm$ 0.06 & 5687 $\pm$ 25.2 & CaT\\
UCD7    & 12.5 & 22.04 $\pm$ 0.02 & 1.41 $\pm$ 0.04 & 5272 $\pm$  15.3 & CaT\\
UCD9    & 14.4 to 24.3 & 22.85 $\pm$ 0.02 & 1.26 $\pm$ 0.05 & 9955 $\pm$ 12.1 & CaT\\
UCD13  & 85.  & 20.97 $\pm$ 0.01 & 1.37 $\pm$ 0.03 & 5292 $\pm$  13.7 & CaT\\
UCD17  & 18.4 & 22.97 $\pm$ 0.03 & 0.28 $\pm$ 0.04 & 5286 $\pm$ 17.3 & H$\alpha$ (emission)\\
UCD19  & 13.1 & 21.88 $\pm$ 0.02 & 1.17 $\pm$ 0.03 & 5680 $\pm$ 11.1  & CaT \\
UCD21  & 31.2 & 22.23 $\pm$ 0.02 & 1.10 $\pm$ 0.04 & 5668 $\pm$   10.9 & CaT\\
UCD23  & 11.5 & 22.36 $\pm$ 0.02 & 1.14 $\pm$ 0.05 & 5693 $\pm$ 14.5 & CaT\\
UCD25  & 11.5 & 21.66 $\pm$ 0.01 & 1.54 $\pm$ 0.03 & 4750 $\pm$ 14.5 & CaT \\
UCD29  & 22.4 & 22.80 $\pm$ 0.02 & 0.54 $\pm$ 0.04 & 5250 $\pm$ 16.5 & H$\alpha$ (emission)\\
UCD40  & 13.9 & 22.12 $\pm$ 0.02 & 1.47 $\pm$ 0.04 & 5632 $\pm$ 10.2 & CaT \\
\hline
Object & R$_{e}$ & I & (V--I)$_{0}$ & $v$ & line used \\
& (pc) & (mag) & (mag) & (km~s$^{-1}$) & \\
\hline
UCD60  & 49.0 & 21.43 $\pm$ 0.02 & 0.96 $\pm$ 0.04 & 3735  $\pm$ 15.8 & CaT \\
UCD65  & 13.6 & 22.05 $\pm$ 0.02 & 1.05 $\pm$ 0.05 & 7681  $\pm$ 15.3 & CaT\\
\hline
\end{tabular}
\end{center}
\medskip
*The effective radius for UCD13 is the result of a single S\'ersic fit to its light distribution. All other sizes are the result of King profile fits.

\label{ucddat}
\end{table*}%

\section{The Perseus Cluster UCD population}
\label{perucd}

\begin{figure}
\begin{center}
\includegraphics[width=0.48\textwidth]{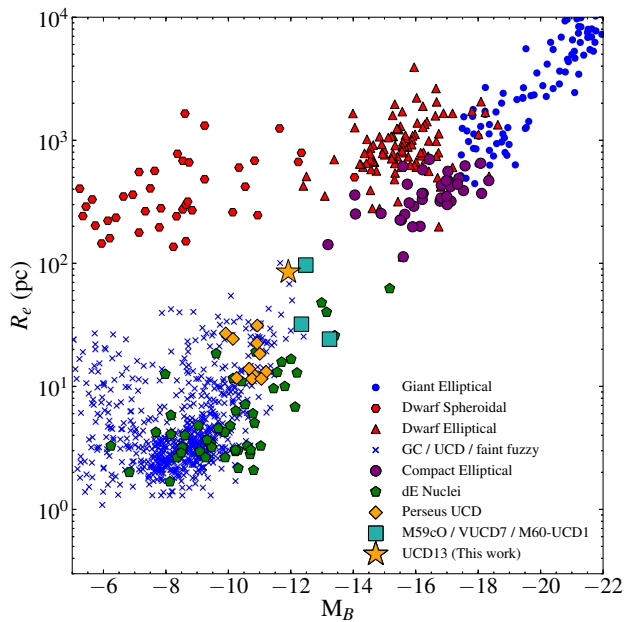}
\caption{Size luminosity distribution for fourteen UCDs in Perseus with confirmed cluster membership compared to old-aged, pressure supported systems. Perseus UCDs are shown as orange rectangles, with UCD13 shown as an orange star. The data sources are identical to Brodie et al. (2011) and Forbes et al. (2013), and are listed in Section~\ref{perucd}. }\label{sizemag}
\end{center}
\end{figure}

Of the 14 UCD candidates with good spectra we are able to confirm 13 of them as belonging to the Perseus cluster (mean velocity of 5366~km~s$^{-1}$), with one (UCD9) likely located just outside of the main cluster.
This implies a background contamination rate in our HST selected UCD candidates (Penny et al. 2012) of $\le$7\%. 
The candidate UCD9 has a measured velocity of $9955 \pm 40.5$~km~s$^{-1}$. Relative to the velocity dispersion of the elliptical galaxies in the cluster (i.e. 1324~km~s$^{-1}$) it is 3.5$\sigma$ from the mean. 
However, relative to the dwarf galaxies (1818~km~s$^{-1}$) it is only 2.5$\sigma$ (see Penny et al. 2008). 
Thus we conclude that UCD9 is likely associated with the large-scale filamentary structure / infall regions surrounding the Perseus cluster, and include it in our sample of Perseus UCDs. We take the size of $R_{e} = 14.4$~pc for UCD9 presented in Penny et al. (2012) as a lower limit on UCD9's effective radius. We calculate an upper limit for the UCDs size if it is instead background to Perseus using the online cosmology calculator of \citet{2006PASP..118.1711W}, assuming H$_{0} = 71$~km~s$^{-1}$~Mpc$^{-1}$. At $z=0.033$ (the redshift of UCD9), the angular scale is 590~pc~arcsec$^{-1}$, vs  350~pc~arcsec$^{-1}$ for Perseus. 
The upper limit on the size of UCD9 is $R_{e} = 24.3$~pc, compared to the size $R_{e} = 14.4$~pc presented in \citet{Penny12} calculated assuming the UCD is located in the Perseus Cluster. Both the upper and lower limit on UCD9's size are presented in Table~\ref{ucddat}. 

The effective radii of the Perseus UCDs were calculated in Penny et al. (2012) by converting the FWHM of their King profile \citep{1962AJ.....67..471K} to an effective radius using the conversion $R_{e} = 1.48 \times \rm{FWHM}$. 
While King profiles are generally a good fit to the light profiles of globular clusters and small UCDs, they are less robust for more extended UCDs. 
Of the UCDs in the Penny et al. (2012) sample, one object, UCD13 is extended enough that we are able to measure its $R_{e}$ directly with a S\'ersic fit to its light distribution (Section~\ref{surf}).

In Figure~\ref{sizemag} we show the size luminosity distribution for all 15 confirmed Perseus UCDs, along with confirmed old-aged ($>5$~Gyr) pressure-supported systems from the compilation of Brodie et al. (2011), which was updated by Forbes et al. (2013). The data presented in Figure~\ref{sizemag} are taken from the following sources: \citet{Barmby2007, Brasseur2011, Brodie2002, 2009Sci...326.1379C, Cockcroft2011, DaCosta2009, Fadely2011, 2013MNRAS.435L...6F,  Foster2011, Geha2003, Harris2010, Hasegan2005, Hasegan2007, Hau09, Hwang2011, Huxor11, Martin2008, Mieske2007, Misgeld2011a, Mouchine2010, 2011MNRAS.414..739N, Richtler2005, Rejkuba2007, Romanowsky2009, SmithCastelli2008, Strader2011, vandenBergh2004}. For a full description of the data compilation used here, see \citet{Misgeld2011,Brodie2011,2013MNRAS.435L...6F}, and references therein.

Where necessary, values of $M_{V}$ were converted to $M_{B}$ using the colour ($B-V$)$=0.96$~mag for a typical elliptical galaxy taken from \citet{1995PASP..107..945F}.  This colour transformation is comparable to the ($B-V$) colours of UCDs in the Fornax Cluster \citep{Karick2003}, and Coma Cluster \citep{2011ApJ...737...86C}, which exhibit colours ranging between ($B-V$)~$=0.65$~mag and ($B-V$)~$=1.1$~mag.

In general, the sizes of the confirmed Perseus UCDs agree with those of UCDs in other studies. 
Furthermore, the sizes of the smaller UCDs with $R_{e} < 20$~pc and $M_{B} \sim -10.5$~mag are in good agreement with those of the \citet{Cote06} sample of dE nuclei. 
This suggests that a fraction of these objects could be the nuclei of stripped dwarf ellipticals, that have had their fragile stellar envelopes removed by the tidal potential of the Perseus Cluster, as well as during encounters with other cluster members. 
These smaller UCDs also overlap in size-luminosity space with the most extended globular clusters, so multiple origins are possible for these UCDs. 
The 14 confirmed UCDs cover a range of colours. 
Thus as well as confirming the status of several old stellar population UCDs, we have confirmed that two of the very blue UCDs (UCD17 and UCD29) with ($B-R$)$<0.6$~mag are associated with the Perseus cluster and must be young massive star clusters.

\begin{figure}
\begin{center}
\includegraphics[width=0.48\textwidth]{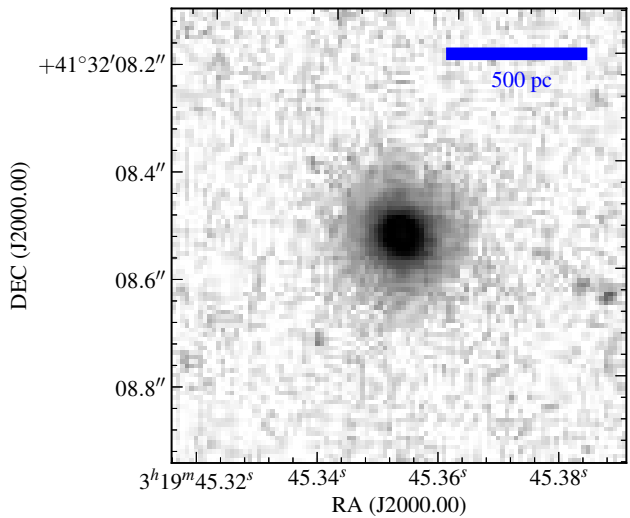}
\caption{$F625W$ ($R$) band HST ACS image of UCD13. A low surface brightness stellar envelope is clearly seen around the UCD.}\label{ucd13im}
\end{center}
\end{figure}

The largest Perseus UCD, UCD13, is approaching the region of the size luminosity diagram occupied by compact ellipticals, and the object exhibits a low surface brightness stellar envelope (Fig.~\ref{ucd13im}). 
An effective radius of 85~pc makes it one of the most extended UCDs identified to date, more extended than a typical globular cluster by a factor of 30. 
At $M_{B} = -11.92$~mag, it is furthermore the brightest UCD with confirmed Perseus Cluster membership.
Given its large size and brightness, UCD13 is an interesting object in which to examine the origin of extended UCDs; its origin as a massive, extended star cluster is more difficult to explain than the smaller Perseus UCDs with effective radii $<20$~pc. We now go on to examine UCD13 in more detail by considering its light profile, kinematics, and stellar population. 

\section{UCD13: Analysis}
\label{ucd13}

\subsection{Surface brightness profile}\label{surf}

\begin{figure*}
\begin{center}
\includegraphics[width=0.48\textwidth]{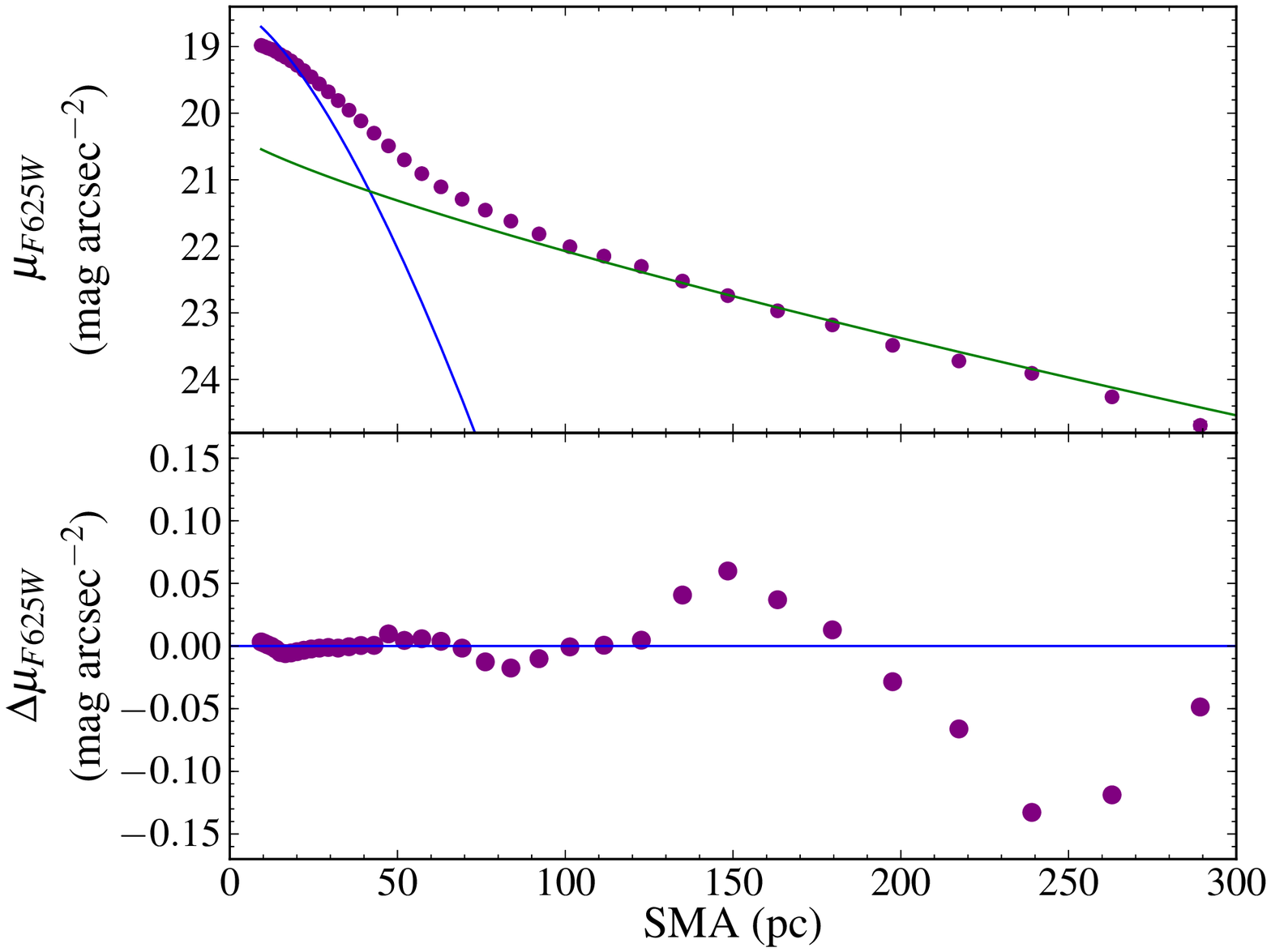}\includegraphics[width=0.48\textwidth]{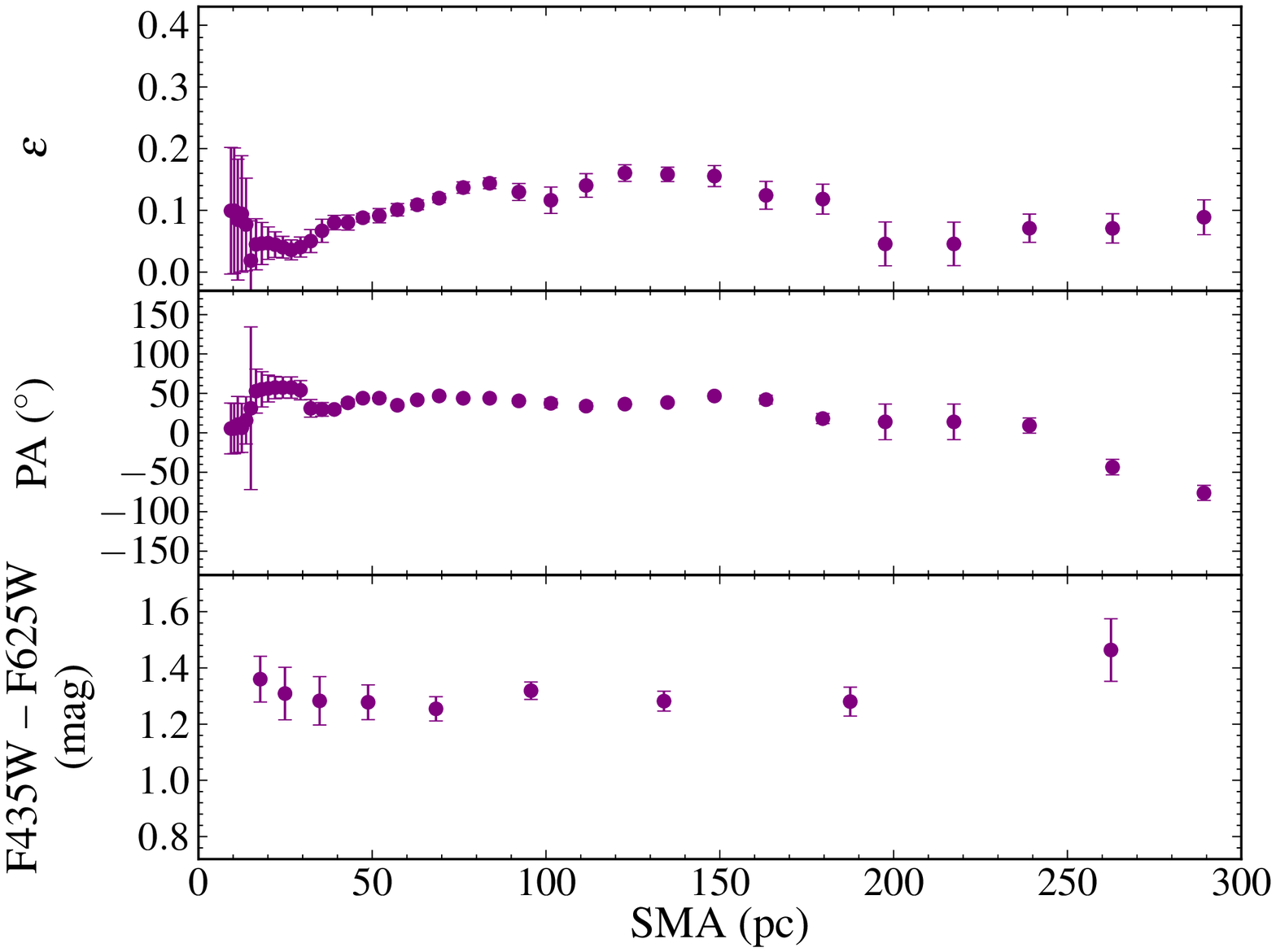}
\caption{An isophotal analysis of UCD13 using HST ACS F625W ($R$) band imaging. The top left-hand plot show S\'ersic profile fits to the surface brightness profile of UCD13 (top), and the bottom left-hand plot shows the residuals between the UCD surface brightness profile and the model profile after convolution with the PSF. The right-hand plot shows the ellipticity (top), position angle (middle), and colour profile for UCD13. The UCD is round, with an average ellipticity $\sim0.1$, and no evidence for isophotal twists. Within the error bars, UCD has a flat colour profile.}\label{allfits}
\end{center}
\end{figure*}

Surface photometry for UCD13 is fit to HST ACS F625W  ($\sim R$) band imaging using the \textsc{iraf} task \textsc{isophote ellipse}.  \textsc{iraf ellipse} fits elliptical isophotes to a galaxy image at predefined semi-major axis values. 
The isophote centres, ellipticities, and position angles are allowed to vary during the fit, with the result of this fit shown in Figure~\ref{allfits}. 
The UCD is remarkably round at all radii, with ellipticity $\epsilon < 0.14$ out to a radius of 280~pc. 
The position angle profile for the UCD does not exhibit any sharp twists that would indicate internal substructure such as spiral arms or a bar, but instead gradually varies out to the last reliable isophote at a semimajor axis of 280~pc. 
The UCD's F435W-F625W ($B-R$) colour profile is flat within the error bars, suggesting the UCD does not exhibit a metallicity gradient.

We model UCD13's light distribution with \textsc{galfit} \citep{2002AJ....124..266P}, using the form of the S\'ersic profile as defined in \citet{Caon93}:

\begin{equation}
\mu(R) = \mu_{e} + \frac{2.5b_{n}}{\text{ln}(10)}[(R/R_{e})^{1/n} - 1],
\end{equation}

\noindent where $R$ is the semi-major axis, $R_{e}$ is the effective radius, and $n$ is the S\'ersic index.  The constant $b_{n}$ is defined in terms of the S\'ersic index as  $b_{n} = 1.9992n - 0.3271$ for the range $0.5 < n < 10$.

The analytic profile for the UCD is convolved with the HST ACS point spread function (PSF) prior to the modelling of the UCD. The PSF is constructed using \textsc{tiny tim} \citep{1993ASPC...52..536K} at the position of UCD13 on the ACS detector to take into account the distortions that are present in the original image. \textsc{galfit} then constructs a model of the UCD that best fits the data by minimising the residuals between the model and the original input image.  

We first fit UCD13 with a S\'ersic model. A single S\'ersic profile fit has $R_{e} = 85\pm 1.1$~pc, $n=2.99\pm0.03$, and $\mu_{F625W} = 21.59\pm0.02$~mag~arcsec$^{-2}$. However, the single S\'ersic profile is a poor fit to the data, with large residuals between the measured and model surface brightness ($\Delta\mu_{F625W} > 0.1$~mag~arcsec$^{-2}$), especially at large radii. UCDs are commonly found to be best fit with two component models (e.g. \citealt{Chilingarian11}), so we investigate if this is the case for UCD13. 

For a two component fit, the inner profile has $R_{e} =  22\pm0.2$~pc, $n=0.67\pm0.02$, $\varepsilon=0.90\pm0.01$, and $\mu_{F625W} = 19.5\pm0.01$~mag~arcsec$^{-2}$. The outer component has $R_{e} = 144\pm 2.1$~pc, $n=1.28\pm0.01$, $\varepsilon=0.90\pm0.01$, and $\mu_{F625W} = 22.67\pm 0.02$~mag~arcsec$^{-2}$. The results of this fit, along with the residuals, are shown in the left-hand plot of Figure~\ref{allfits}. UCD13 is well fit by a two-component S\'ersic model, with typical residuals $\Delta\mu_{F625W} < 0.05$ between the surface brightness profile of UCD13 and the PSF-convolved model of UCD13 constructed using \textsc{galfit}.

\subsection{Internal kinematics}
\label{sec:cE}

\begin{figure}
\begin{center}
\includegraphics[width=0.48\textwidth]{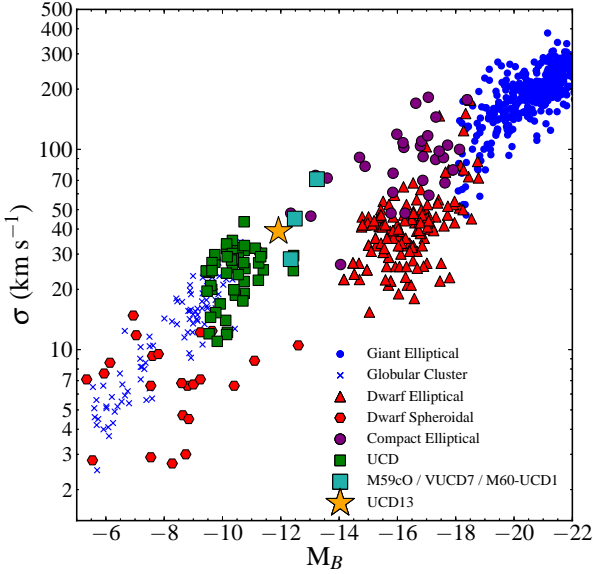}
\caption{The velocity dispersion-luminosity relation $\sigma$ vs. $M_{B}$ for  early type systems. The data sources are identical to \citet{2011MNRAS.413.2665F} and references therein. UCD13 is the only Perseus UCD with a measured velocity dispersion $\sigma_{0}$, and is represented by an orange star. UCDs from other groups and clusters are shown as green squares, unless otherwise denoted. UCD13 lies at the bright end of the relation seen for Ultra Compact Dwarfs and UCDs, with dEs and dSphs having lower values of $\sigma$ at a comparable magnitude. Its central velocity dispersion $\sigma_{0} = 38$~km~s$^{-1}$ is comparable to that of a dE of $M_{B}=-16$~mag.}\label{msig} 
\end{center}
\end{figure}

\begin{table*}
\caption{Basic properties of UCD13. Also shown for comparison is a compact elliptical in the Virgo Cluster (M59cO, \citealt{Chilingarian08b}), along with VUCD7 \citep{2007AJ....133.1722E}, a UCD in the Virgo Cluster  with similar properties to UCD13. M60-UCD1 from \citet{2013ApJ...775L...6S}, the densest galaxy identified to date, is also listed.}\label{props}
\begin{center}
\begin{tabular}{lcccc}
\hline
Property & UCD13 & M59cO & VUCD7 & M60-UCD1\\
\hline
$\alpha$  (J2000.0) & 03:19:45.13 & 12:41:55.3 &  12:31:52.93 & 12:43:36.98\\
$\delta$ (J2000.0) & +41:32:06.0 & +11:40:04 & +12:15:59.5 & +11:32:04.92\\
$M_{B}$ (mag) & --11.92 & --12.34 & --12.48* & --13.24\\
R$_{e}$ (pc) & $85 \pm 1.1$ & $32$ & 96.8 & $24.2 \pm  0.5$\\
S\'ersic $n$ & $2.99 \pm  0.38$ & 1.0 & 2.2 & $3.53 \pm 0.01$ \\
$\sigma_{0}$ (km~s$^{-1}$)& $38\pm9$ & $28.3 \pm 2.9$** & $45.1 \pm 1.5$ & $71 \pm 5$ \\
$M_{\star}$  (M$_{\odot}$) & $(4.4 \pm 0.9)\times10^{7}$ & $9.1\times10^{7}$  & $(8.8\pm2.1)\times10^{7}$ & $2.4\times10^{8}$\\
$M_{dyn}$  (M$_{\odot}$)& $(2.3 \pm 0.8) \times 10^{8}$ & $8.1\times10^{7}$ & $(1.62\pm5.7)\times10^{8}$ & $(2.0 \pm 0.3) \times 10^{8}$ \\
$[Z/H]$ (dex)& $-0.52^{+0.33}_{-0.29}$  & $-0.03 \pm 0.04$ & --1.35 to --0.33 & $+0.26 \pm 0.02$ \\
\hline
\end{tabular}
\end{center}
\medskip
*Converted from $M_{V} = -13.44$ assuming $(B-V) = 0.96$, typical for early-type systems \citep{1995PASP..107..945F}. **Value taken from Norris et al., in prep.
\end{table*}

To help establish if this object is the stripped remnant of a much larger system, we measure its internal velocity dispersion using the Penalised Pixel Fitting (\textsc{ppxf}) routine (Cappellari \& Emsellem, 2004).  
The instrumental velocity dispersion is confirmed using two stellar spectra that were taken as serendipitous observations during the DEIMOS observations (these stars fell into the slits of our target objects). 

We measure UCD13 to have an internal velocity dispersion  $\sigma = 35\pm8$~km~s$^{-1}$, determined from the CaT~1 and CaT~2 lines, and confirmed using the H$\alpha$ line. The CaT~3 line was affected by a strong sky-line, so was not used in the determination of $\sigma$. We then correct this global internal velocity dispersion $\sigma$ to a central velocity dispersion $\sigma_{0}$ (the velocity dispersions within one $R_{e}$) \citep{}. Using an effective radius $R_{e} = 85$~pc as found for a single profile fit, we obtain  $\sigma_{0} = 38 \pm 9$~km~s$^{-1}$. 

We calculate the dynamical mass of the UCD using the virial theorem. To calculate the total mass of the object, we assume an exponential surface brightness profile ($n=1$) as a compromise between the inner and outer profiles, and a half-light radius $R_{e} = 85$~pc. Using the virial theorem with a correction for the S\'ersic profile shape provided in \citet{2004PASP..116..138C}, this gives us a dynamical mass for the object of $(2.3 \pm 0.8) \times 10^{8}$~M$_{\odot}$.

We estimate a stellar mass-to-light ratio ($M/L$) using the stellar population synthesis models of \citet{Maraston2005}. We assume a Salpeter IMF, an old stellar population with an age of 10~Gyr, and a metallicity [Z/H]~$=-0.33$~dex, comparable to the UCD's metallicity presented in Section~\ref{metal}. Based on this, we find a $B$-band $M/L$ of $5.75$. The absolute $B$-band magnitude of UCD13 ($M_{B} = -11.92$~mag) is converted to a solar luminosity and then multiplied by the stellar $M/L$, giving a stellar mass for the object of $(5.1 \pm 0.9)\times10^{7}$~M$_{\sun}$. This gives  $M_{dyn}/M_{\star}$ of 4.5. However, the IMF for UCDs may be better described by a Kroupa IMF (e.g. \citealt{Chilingarian11}). Using a Kroupa IMF with [Z/H]$=-0.33$~dex and an old stellar population, the $B$-band stellar mass-to-light ratio will decrease to 3.66, resulting in $M_{dyn}/M_{\star}$ = 7.2. Both results for $M_{dyn}/M_{\star}$ are inconsistent with a highly dark matter dominated object. 

%M/L = 5.75 Kroupa

\subsection{Metallicity}
\label{metal}

We use the method of \citet{2012MNRAS.426.1475U} to measure the metallicity of UCD13 from the strength of the near infrared calcium triplet spectral feature. Sensitive to metallicity and insensitive to ages older than 3 Gyr \citep{2003MNRAS.340.1317V}, the strength of CaT is independent of $\alpha$-element enhancement and horizontal branch morphology \citep{2012ApJ...759L..33B}. However the strength of the CaT is dependent on the IMF as the CaT is weaker in a stellar population with a bottom heavy IMF, such as those found in massive early type galaxies by \citet{2010Natur.468..940V}, than in a population with a Kroupa IMF and the same metallicity \citep{2003MNRAS.340.1317V}.

To remove the effect of strong sky line residuals on the CaT measurement, we mask the sky line regions and fit a linear combination of stellar templates to the observed spectrum.
We then measure the strength of CaT on the fitted spectrum and use a Monte Carlo resampling technique to estimate a confidence interval on the measured line strength.
Lastly we use the single stellar population models of \citep{2003MNRAS.340.1317V} to convert the line strength into a metallicity.
See \citet{2012MNRAS.426.1475U} for further details of the metallicity measurement process.

We measure a metallicity of [Z/H] $= -0.52_{-0.29}^{+0.33}$~dex from the strength of the CaT.
Using the GC colour--metallicity relation of \citet{2012MNRAS.426.1475U} we derive a colour based metallicity of [Z/H] = $-0.13 \pm 0.23$~dex, consistent with the CaT based metallicity.
The metallicity and other properties of UCD13 measured in this work are summarised in Table~\ref{props}, along with three comparison objects from the literature (M59cO, \citealt{Chilingarian08b}; VUCD7, \citealt{2007AJ....133.1722E}; and M60-UCD1, \citealt{2013ApJ...775L...6S}).

\section{UCD13: tidally stripped galaxy or star cluster?}
\label{sec:ucd13}

Could UCD13 be the tidal remnant of a once much larger galaxy? UCD13 has a velocity $v=5292 \pm 13.7$~km~s$^{-1}$, compared to $v=5276$~km~s$^{-1}$ for NGC~1275 \citep{2007ApJS..170...33P}.
At an angular separation of $1.5'$, UCD13 and NGC 1275 have a minimum physical separation of 30~kpc. Such conditions are highly favourable for galaxy-galaxy interactions and tidal stripping.  NGC1275 furthermore exhibits a shall structure consistent with the BCG actively disrupting and accreting lower mass galaxies today (e.g. Penny et al., 2011), and UCD13's progenitor likely contributed to these shells. 
It is therefore highly likely the two objects are interacting given their small radial velocity separation of $<20$~km~s$^{-1}$ and close physical proximity, so we examine UCD13 for features that might indicate that it is a UCD formed via the tidal stripping of a low mass galaxy. 

\subsection{The velocity dispersion-luminosity relation}
\label{sec:vlrel}

When a galaxy is tidally stripped, its effective radius and luminosity will be reduced, but its central velocity dispersion and central metallicity will remain largely unchanged from those of its progenitor \citep{1992ApJ...399..462B}. As the tidal remnant relaxes after stripping, its effective radius will increase slightly,  causing its central velocity dispersion to decrease, however this effect will be small. The simulations of \citet{2009Sci...326.1379C} show that for a disc galaxy, the stellar disc of the galaxy will be heavily stripped, and its central velocity $\sigma_{0}$ will decrease by $<5$~km~s$^{-1}$. This decrease in velocity dispersion is within our error bar of $\pm8$~km~s$^{-1}$, and we can assume that the progenitor and remnant galaxies will have the same value of $\sigma_{0}$.

For the tidal stripping of a nucleated dE with a fragile stellar envelope (i.e. most of its mass is located in its central regions), we hypothesise than this decrease in $\sigma_{0}$ will be even smaller. In addition, as shown in Fig.~\ref{msig}, at a given magnitude dEs exhibit a scatter in their measured internal velocity dispersions, with $\sigma_{0}$ ranging from $\sim 20~$km~s$^{-1}$ to $70$~km~s$^{-1}$.  Any decrease in $\sigma_{0}$ resulting from the tidal transformation of a dE to a UCD given our measured value of $\sigma_{0} = 38$~km~s$^{-1}$ will fall within this range. We can therefore use our measurement of UCD13's central velocity dispersion to characterise the nature of its progenitor.

Old, pressure supported galaxies follow a well-defined relation between $\sigma$ and luminosity: the Faber-Jackson relation (Faber \& Jackson 1976). 
However, the slope of this relation is observed to change in the dwarf elliptical regime, from $L \propto \sigma^{4}$ for giant ellipticals, to $L \propto \sigma^{2}$ for dEs  \citep{Matkovic05}. This  allows us to more easily separate the kinematics of giants and dwarf ellipticals. 
We can therefore use the luminosity-sigma relation to estimate the mass of the progenitor of UCD13 by examining its location on the velocity dispersion--luminosity relation (Fig.~\ref{msig}), with a high value of $\sigma_{0}$ likely indicative of a tidal stripping origin.

UCD13's central velocity dispersion  is comparable to the internal velocity dispersion of a dE galaxy with $M_{B} \sim -16$~mag ($\sigma_{0} = 38$~km~s$^{-1}$, Fig.~\ref{msig}). 
We can rule out UCD13 as currently being a dSph with a high surface brightness core, as its velocity dispersion is higher than that of dSph galaxies of comparable magnitude. In the Local Group, Fornax, Sculptor and Leo~I have have luminosities comparable to UCD13 ($-12~{\rm mag} < M_{V} < -10~{\rm mag} $), but  central velocity dispersions $\sigma < 12$~km~s$^{-1}$ \citep{Gilmore07,Walker07}. 
The UCD's $\sigma_{0}$ is too low for it to be the remnant of a low-mass elliptical galaxy or bulge. 
We confirm the luminosity of UCD13's progenitor by utilising the relations of \citet{Matkovic05}.
Using the relations presented in their figures 4 and 5, $\sigma_{0}~=~38$~km~s$^{-1}$ corresponds to a galaxy with $M_{R}=-17.5$~mag and $(B-R)~=~1.35$~mag. UCD13 has $(B-R)_{0}~=~1.37$~mag, in strong agreement with this predicted progenitor colour. 
Therefore, based on colour and internal velocity dispersion, the progenitor of UCD13 is likely a dwarf elliptical galaxy with $M_{B}~=~-16$~mag. 

We should note than an internal velocity dispersion of  $\sim40$~km~s$^{-1}$ is not inconsistent with  the velocity dispersion-magnitude relation for massive star clusters.
However, the size of UCD13 ($R_{e}=93$~pc) is higher than that expected for an extended star cluster with a comparable $\sigma_{0}$. 
Despite having a higher velocity dispersion of $\sigma = 45 \pm 5$~km~s$^{-1}$, the massive luminous star cluster W3 has a size $R_{e} = 17.5\pm1.8$~pc \citep{2004A&A...416..467M}, one-fifth the size of UCD13. 
For a comparable central velocity dispersion, massive star clusters are more compact than UCD13. A massive star cluster origin for UCD13 is examined in detail in Section~\ref{starcluster}.

\subsection{Light distribution}

The surface brightness distributions of dEs are generally well fitted by an exponential $n\approx 1$ S\'ersic profile, while the light distributions of giant ellipticals closely follow a de Vaucouleurs profile with a S\'ersic index $n\sim4$. 
Although the luminosities of cEs frequently overlap those of dEs ($M_{B} > -18$), the compact ellipticals are easily separated from dEs due to their concentrated light distributions from a single S\'ersic profile fit, with $n >2$. 
With $n=2.99$ from a single profile S\'ersic fit, UCD13 has a concentrated light distribution resembling that of a typical compact elliptical galaxy.

To establish if the object's large size and luminosity could be the result of the tidal stripping of a dE, in Fig~\ref{sizemag} we include for comparison a sample of dE nuclei from \citet{Cote06}. UCD13 is more extended than a dE nucleus of comparable magnitude by a factor of 10. \citet{Evstigneeva08} find a comparable result for UCDs in the Fornax and Virgo clusters, with UCDs having effective radii larger than $20$~pc, twice the size of those of early-type dE nuclei at the same luminosity. 

However, during a tidal stripping event, the nucleus may be able to retain or re-accrete some of its envelope of tidally stripped stars. 
If a stripped dE is able to retain some of its stellar envelope in addition to its nucleus, the resulting UCD will be more extended than a dE nucleus alone. 
\citet{2013MNRAS.433.1997P} show that UCDs with sizes $R_{e} > 10$~pc can be formed from the tidal stripping of a nucleated dE if the dE has a small pericentric approach ($<10$~kpc) from the perturbing galaxy. 
The dE must be on a highly elliptical orbit, and must be on its first or second approach for such a transformation to occur. 
In their simulations, this scenario produces extended UCDs with sizes $\sim100$~pc. 
The resulting UCDs consist not only of the remnant dE nucleus, but also some of the dE's retained stellar envelope. 

Given its close proximity to NGC 1275 ($\sim30$~kpc) and small separation in radial velocity ($<20$~km~s$^{-1}$), UCD13 would currently be near the apocentre of a radial orbit, else we are observing its orbit face-on (such an orbit could still be a radial). 
If UCD13 is indeed at an orbit apocentre $\sim30$~kpc after several pericentric approaches to NGC 1275, dynamical friction will have decayed its original highly elliptical orbit into a circular one, as predicted in the simulations of \citet{Chilingarian11}. 
Therefore a dE stripping scenario is a feasible explanation for its origin. 

\subsection{The mass-metallicity relation}

Assuming a dE rather than a massive cluster origin for UCD13, we proceed to examine the mass-metallicity relation for dwarf galaxies to estimate the mass of the UCD's progenitor. The mass-metallicity relation for galaxies presented in figure~6 of \citet{2009MNRAS.396.2103M} shows that a galaxy with [Z/H]$ = -0.52$~dex has a mass $\sim 10^{9}$~M$_{\sun}$. This mass is comparable to that of a dE with $M_{B} \sim -16$, in agreement with the result using the Faber-Jackson relation presented in Section~\ref{sec:vlrel}.

\section{A massive star cluster?}
\label{starcluster}

A galaxy-like central bulge with a stellar envelope as exhibited by UCD13 (Fig.~\ref{ucd13}) does not necessarily exclude a massive star cluster origin. 
The simulations of \citet{2005MNRAS.359..223F} showed that UCDs with diffuse stellar envelopes similar to that observed for UCD13 can be formed when massive ($>10^{7}$~M$_{\odot}$) star cluster complexes merge. 
Such objects will have circular isophotes in their inner regions, with their ellipticities increasing at large radii.  
\citet{2011MNRAS.414..739N} suggested $M_{V} = -13$~mag as the upper magnitude limit for UCDs formed as giant globular clusters, with UCD13 approaching this magnitude limit assuming $(B-V) \sim 1$, a typical value for UCDs (e.g. \citealt{2011ApJ...737...86C}).  
Multiple mergers between multiple massive star clusters would be required to reach this size, and the simulations of \citet{2002MNRAS.330..642F,2005MNRAS.359..223F} are able to form UCDs via this method. 

UCDs formed  out of a merged star cluster complex would have a core of merged star clusters, with an envelope of stripped stars lost during the merger of the star clusters- similar in morphology to UCD13. 
The intermediate age star cluster W3 ($R_{e} = 17.5$~pc) in the merger remnant NGC~7252 is likely a UCD-like object \citep{2005MNRAS.359..223F} recently formed via this method. 
However, its size is only one-quarter that of UCD13, so it is unclear if massive star clusters such as W3 are the same class of objects as the most extended UCDs. 
We therefore examine the sizes of massive star clusters surrounding NGC~1275 (e.g. \citealt{2010MNRAS.405..115C,Penny12}) to address this question. 

NGC~1275 is surrounded by a complex system of filaments, a number of which host regions of active star formation. 
These star forming regions are hypothesised as a region of UCD formation, with several proto-UCDs located in these filaments exhibiting $(B-R)_{0}$ colours $<0.55$, implying ages $<100$~Myr. 
However, the sizes of these young UCD-like star clusters are far smaller than that found for UCD13, with the largest blue UCD, UCD29, having a size $R_{e} = 22.4$~pc \citep{Penny12}. 
Instead, objects such as UCD13 are more likely to form in lower density regions of the cluster. UCD13 and NGC~1275 are very close in both radial velocity and angular separation, implying UCD13 is located in the centre of Perseus, a region of the cluster that is not conducive to mergers between massive star clusters. 
A strong tidal field will prohibit merger events between massive star clusters.  
\citet{2010MNRAS.405..115C} showed that a number of young, massive star clusters in the star forming filaments surrounding NGC~1275 are being tidally stripped and will eventually fall into the galaxy's nucleus, and thus NGC~1275's filaments may not be an ideal location for forming more extended UCDs via star cluster mergers.  

Mergers between multiple massive young, UCD-sized star clusters are required to produce an extended object such as UCD13. 
A star-cluster merger formation scenario for UCD13 therefore seems unlikely. However, other Perseus UCDs with sizes $<20$~pc are likely a mix of massive globular clusters and galaxy nuclei. 

\section{Discussion}
\label{sec:discuss}

The UCD population in Perseus may have multiple origins. UCDs with smaller effective radii are likely an extension of the globular cluster system to higher magnitudes and sizes. 
However, the origin of UCDs more massive than this cannot be explained simply as massive globular clusters. 
Multiple mergers between massive globular clusters or star forming complexes are required to explain their origin. 
UCDs with galaxy-like properties and cEs likely represent a family of objects with a tidal stripping origin, with compact ellipticals similar to the Local Group object M32 being formed from the tidal stripping of low-luminosity spiral galaxies \citep{2001ApJ...557L..39B}, while the most extended UCDs with sizes $>40$~pc are probably stripped dEs. 
Massive UCDs would therefore simply be an extension of cE galaxies to lower masses and sizes into a regime where they overlap with the most massive globular clusters

Based on its size and internal velocity dispersion, we have established that UCD13 is likely the tidal remnant of a larger progenitor. 
Similar objects to UCD13 are shown in Table~\ref{props}, and include M59cO \citep{Chilingarian08b} at $M_{B} = -12.34$~mag with $R_{e} = 32$~pc, located at a distance of 9~kpc from the Virgo Cluster giant elliptical M59. 
This object has a blue core surrounded by a redder envelope, with an age $9.3 \pm 1.4$~Gyr, and a metallicity of $Z = -0.03 \pm 0.04$~dex. 
Its internal velocity dispersion is $28.3$~km~s$^{-1}$ (Norris et al., in prep), and lies between the UCDs and cEs on the fundamental plane. 
\citet{Chilingarian08b} suggested it is similar in metallicity to low-luminosity ellipticals and lenticulars in terms of age, metallicity, and effective surface brightness.  
UCD13 and M59cO have very similar properties, with M59cO exhibiting a two component structure with a low surface brightness stellar envelope. 

\citet{Chilingarian08b} suggested M59cO is the result of tidal stripping by M59, with the UCD/cE transition object the remnant of a much larger object. \citet{2013ApJ...775L...6S} reach a similar conclusion for the origin of M60-UCD1, which is likely the densest galaxy identified to date with a mass comparable to UCD13, but a radius $R_{e} = 24$~pc.
Further examples of UCDs in Virgo with properties comparable to UCD13 with $\sigma_{0} \sim 30$~km~s$^{-1}$ are presented in \citep{Evstigneeva08}. VUCD~7 from their sample is almost identical to UCD13, with $\sigma_{0} = 45.1$~km~s$^{-1}$, M$_{V} = -13.44$~mag, and $R_{e} = 96.8$~pc. Further examples of extended UCDs are also found in the Coma Cluster with $M_{B} \sim -13$~mag \citep{2009MNRAS.397.1816P}. 
Therefore objects like UCD13 likely represent an extension of cEs to lower masses, forming via the tidal stripping of dEs rather than low mass disk galaxies.

\section{Conclusions}
\label{sec:conclude}

We have examined the properties of Perseus UCDs with their cluster membership confirmed via \textsc{keck deimos} spectroscopy. 
Perseus UCDs with sizes $<20$~pc are consistent with being massive globular clusters or the nuclei of stripped dwarf ellipticals. Two blue UCDs with colours ($B-R$)~$<0.6$ are confirmed as members of the Perseus Cluster, and must have formed within the last 100~Myr as massive star clusters in the star forming filaments surrounding NGC 1275. 

The largest UCD (UCD13) is likely the nuclear remnant of a tidally stripped dwarf elliptical that has retained a large fraction of its stellar envelope, consistent with the simulations of \citet{2013MNRAS.433.1997P}. Its dynamical mass is $(2.3 \pm 0.8)\times10^{8}$~M$_{\sun}$, making it one of the most massive UCDs identified to date. 
Its internal velocity dispersion, $\sigma_{0} = 38 \pm 9$~km~s$^{-1}$, along with its metallicity $\rm{[Z/H]} = -0.52^{+0.33}_{-0.29}$~dex, and colour $(B-R)_{0} = 1.37$, are consistent with its origin as a dE with an $M_{B}$ of $-16$ and a mass $\sim10^{9}$~M$_{\odot}$. 
Its size ($R_{e} = 85 \pm 1.1$~pc) and location at the centre of Perseus, the UCD is inconsistent with being a massive star cluster. 

\section*{Acknowledgments}

SJP acknowledges the support of an Australian Research Council Super Science Postdoctoral Fellowship grant FS110200047. DAF thanks the ARC for support via DP130100388. This work was supported by NSF grant AST-1109878. The data presented herein were obtained at the W.M. Keck Observatory, which is operated as a scientific partnership among the California Institute of Technology, the University of California and the National Aeronautics and Space Administration. The Observatory was made possible by the generous financial support of the W.M. Keck Foundation. The authors wish to recognise and acknowledge the very significant cultural role and reverence that the summit of Mauna Kea has always had within the indigenous Hawaiian community. We are most fortunate to have the opportunity to conduct observations from this mountain. We thank Jacob Arnold and Lee Spitler who carried out the Keck-DEIMOS observations presented in this work. We thank Michael J.I. Brown, Nicola Pastorello and Caroline Foster for their assistance with this work.

\end{document}